\documentclass[letterpaper,journal]{IEEEtran}

\usepackage{amsmath,amssymb,amsfonts}
\usepackage{amsthm}
\usepackage{mathrsfs}
\usepackage{graphicx}
\usepackage{multirow}
\usepackage{booktabs}
\usepackage{array}
\usepackage{algorithm}
\usepackage{algpseudocode}
\usepackage{textcomp}
\usepackage{xcolor}
\usepackage{enumitem}
\usepackage{cite}
\usepackage{xurl}
\usepackage[hidelinks]{hyperref}

\renewcommand{\Comment}[1]{// #1}

\theoremstyle{plain}

\theoremstyle{remark}

\theoremstyle{definition}

\begin{document}

\title{Generalizing Unit Commitment Problem Solving via SAT-based Decoupling}


\author{Yuxin Zhao, Han Huang, Fangji Fu, and Zhifeng Hao%
\thanks{Yuxin Zhao and Fangji Fu are with the School of Software Engineering, South China University of Technology, Guangzhou 510006, China (e-mail: yuxinzhaozyx@163.com; fufangji0202@163.com).}%
\thanks{Han Huang is with the School of Software Engineering, Sun Yat-sen University, Zhuhai 519082, China (e-mail: huangh985@mail.sysu.edu.cn). (Corresponding author: Han Huang.)}%
\thanks{Zhifeng Hao is with the Department of Mathematics, College of Science, Shantou University, Shantou 515063, China (e-mail: haozhifeng@stu.edu.cn).}
\thanks{This work has been submitted to the IEEE for possible publication. Copyright may be transferred without notice, after which this version may no longer be accessible.}
}

\maketitle

\begin{abstract}
As the cornerstone of modern power systems, the Unit Commitment Problem (UC) is critical for ensuring operational security and economic efficiency in the ongoing global energy transition. However, existing UC studies typically propose specialized algorithms for specific variants and operational requirements, tightly coupling the algorithms to their target models and limiting their applicability to other variants. To address this issue, this paper proposes a method that uses SAT-based reduction to decouple the algorithm from the problem, which allows a single algorithm to solve multiple UC variants. By uniformly reducing all UC variants to SAT instances solvable by standard SAT solvers, this method makes the solving algorithm independent of the original UC variant, thus granting it broad applicability across diverse variants. Experimental results show that our method achieves better solution quality than specialized algorithms and demonstrates stronger generalizability. This work offers a fast and flexible framework for addressing newly emerging UC formulations in evolving power systems.
\end{abstract}

\begin{IEEEkeywords}
Unit Commitment Problem, Decoupling, SAT, Reduction.
\end{IEEEkeywords}

\section{Introduction}\label{sec1}

\IEEEPARstart{T}{he} Unit Commitment Problem (UC)\cite{kerrUnitCommitment1966} is one of the core optimization problems in power system. Its goal is to determine the on/off status and power output of each unit over a scheduling time period, satisfying load demand and operational constraints while minimizing the total operating cost of the power system.
With the ongoing global energy transition\cite{Hong:2026kxc,huScalableControlLargeScale2025, yangBurdenHydropowerUnits2018, guoGridIntegrationFeasibility2023, luCombinedSolarPower2021, mccardleCollaborationsDriveEnergy2023,srivastavaNewAggrandizedClass2022,teoOptimizationFuzzyEnergyManagement2021,wangConditionKnowledgeRepresentation2024}, UC now includes not only conventional fossil-fuel units\cite{aliSolvingFuelBasedUnit2024, liuHMAMRLMulticriterionFlexible2026, byeonUnitCommitmentGas2020} but also units of renewable sources such as wind\cite{wangRobustRiskConstrainedUnit2017, duOperationHighRenewable2019}, solar\cite{duOperationHighRenewable2019, zhangMultipleDistributedPVs2025, dongDatadrivenCostBudget2025, jainOptimizedUnitCommitment2025}, and nuclear power\cite{zhangTwostageStochasticUnit2024}. This evolution has significantly increased the complexity of the problem.

To address diverse operational requirements, many variants of UC have emerged. Examples include UC with ramping constraints\cite{liDistributed$Q$LearningBased2020, correa-posadaDynamicRampingModel2017, jinDataDrivenLookAheadUnit2019, liuAccurateModelingConfiguration2020, elsayedThreestagePriorityList2021, faizahDynamicPriorityApproach2024, karabasExactSolutionMethod2023}, security-constrained UC\cite{sunLagrangianDecompositionApproach2016, xavierTransmissionConstraintFiltering2019, ahmadiSecurityConstrainedUnitCommitment2019, gaoInternallyInducedBranchandcut2022,luEvolutionaryFractionalOrderExtended2025,wangScalableAdaptiveFuzzy2026, quLinearizationMethodLargescale2024,zhangIncorporatingFrequencySecurity2026a}, network-constrained UC\cite{longACNetworkConstrainedUnit2025,zhaoDistributedResilientSecondary2025, duHighefficiencyNetworkconstrainedClustered2019, wuAcceleratingNCUCBinary2016}, and UC that integrates hydropower\cite{chenEfficientMILPApproximation2016}, wind power\cite{wangRobustRiskConstrainedUnit2017, duOperationHighRenewable2019} or solar power\cite{duOperationHighRenewable2019,zhangMultipleDistributedPVs2025, jainOptimizedUnitCommitment2025, dongDatadrivenCostBudget2025}.
These variants differ in their mathematical structure, making it challenging to solve them all with a single fixed algorithm. In practice, existing approaches typically analyze the specific model of a given variant and design a tailored algorithm.
However, such algorithms are tightly coupled to the problem. When a new variant appears, minor adjustments to an existing algorithm often fail to meet the new requirements, resulting in limited reusability across different variants.
This repeated cycle of designing and implementing custom algorithm is both labor-intensive and inefficient for handling power scheduling tasks in various scenarios.
If the algorithm can be decoupled from the mathematical model of the problem, a single algorithm could solve multiple UC variants. Such decoupling would eliminate the need to design custom algorithms for each new variant and enable rapid responses to diverse unit commitment demands.

The classical UC was first introduced by Kerr et al\cite{kerrUnitCommitment1966}. Over the past half-century, UC has received continuous attention from both industry and academia, leading to many UC variants.
Although significant research\cite{correa-posadaDynamicRampingModel2017, jinDataDrivenLookAheadUnit2019, liuAccurateModelingConfiguration2020, elsayedThreestagePriorityList2021, faizahDynamicPriorityApproach2024, karabasExactSolutionMethod2023, sunLagrangianDecompositionApproach2016, xavierTransmissionConstraintFiltering2019, ahmadiSecurityConstrainedUnitCommitment2019, gaoInternallyInducedBranchandcut2022, quLinearizationMethodLargescale2024, duHighefficiencyNetworkconstrainedClustered2019, wuAcceleratingNCUCBinary2016} has been conducted on these variants and has yielded promising results, most proposed algorithms are tightly coupled to the specific problem.
For example, for the UC with ramping constraints, Faizah et al.\cite{faizahDynamicPriorityApproach2024} proposed a dynamic priority approach in which ramping constraints are incorporated as both priority criteria and boundary conditions.
Elsayed et al.\cite{elsayedThreestagePriorityList2021} proposed a priority list method that includes a dedicated stage to update the priority list based on ramping constraints.
While embedding ramping constraints directly into the algorithm can improve solution efficiency, this embedding also tightly couples the algorithm to this specific constraint. As a result, such algorithms are difficult to apply to other UC variants.

\begin{figure*}[!t]
\centering
\includegraphics[width=0.86\textwidth]{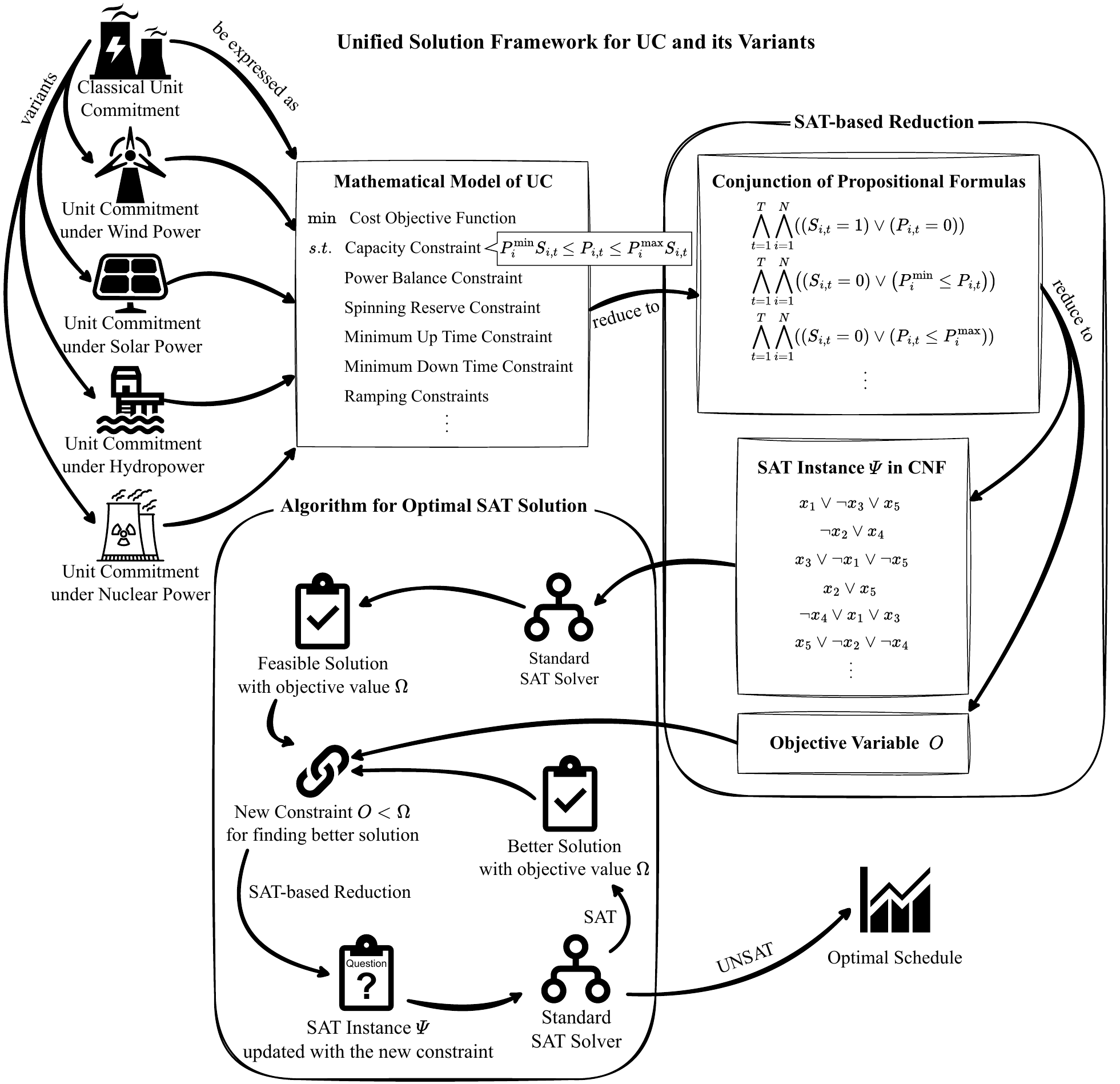}
\caption{Unified Solution Framework for UC and its Variants}\label{fig:UniUC}
\end{figure*}

To address the tight coupling between the algorithm with the mathematical model in UC and its variants, this paper proposes a unified solution framework.
The core idea of this framework is to decouple the algorithm from the problem model through SAT-based reduction.
As shown in Figure~\ref{fig:UniUC}, the framework transforms various UC formulations into Boolean Satisfiability Problems (SAT) and iteratively solves them using a standard SAT solver.
In this framework, the algorithm, i.e. the SAT solver, is completely decoupled from the original mathematical model of the UC. The solver does not require adaption for specific UC variants and can be directly applied to all of them, making the algorithm generalizable across diverse UC variants.

The main contributions of this paper are summarized as follows:
\begin{enumerate}[label=\arabic*)]
  \item We propose a SAT-based reduction method for UC. For generation capacity constraints of UC, we exploit the relationship between the on/off status and output power to equivalently reformulate the constraints, avoiding the introduction of auxiliary variables and multiplication operations. For numerical comparison constraints, we design a binary-comparison-based reduction rule that enables unit propagation to quickly eliminate clauses irrelevant to the current search context. These reduction rules effectively reduce the size of SAT instances and improve solving efficiency.
  \item We propose a SAT-based unified solution framework for UC and its variants. The framework uniformly reduces heterogeneous mathematical model of different UC variants into SAT instances, thereby decoupling the solving algorithm from the specific UC variants and enabling a standard SAT solver to uniformly solve classical UC and multiple UC variants without variant-specific redesign. Experiments on classical UC and it variant with ramping constraints show that the proposed framework achieves better solution quality and stronger generalizability than specialized algorithms.
\end{enumerate}

The remainder of this article is organized as follows. Section~\ref{sec:related-works} reviews existing UC solution methods and discusses their coupling with specific problem formulations. Section~\ref{sec:model} presents the mathematical models of the classical UC and UC with ramping constraints. Section~\ref{sec:method} introduces the proposed unified solution framework for UC and its variants, including the SAT-based reduction and the SAT-based optimization algorithm. Section~\ref{sec:experiments} evaluates the framework on benchmarks for both UC formulations. Finally, Section~\ref{sec:conclusion} concludes the article and outlines directions for future work.

\section{Related Works}\label{sec:related-works}

When faced with new problem variants, traditional algorithms typically exhibit limited generalizability. The fundamental reason lies in the tight coupling between the algorithm and the mathematical model of the problem. Specifically, most existing algorithms are tailored to a specific problem formulation or mathematical structure. When the problem structure changes, the original algorithmic framework cannot be directly applied to the new variant, necessitating redesign or adaptation of the algorithm. This coupling not only limits the generalizability of the algorithm but also increases the time and resource costs required to address emerging problems.

We surveyed algorithmic research on UC and its variants, and identified that the coupling between problem and algorithm can be categorized into two types: (1) algorithmic structure depends on problem structure, and (2) algorithmic parameters depend on problem parameters.

Algorithmic structure depending on problem structure is the most common form of coupling, as many algorithmic researches achieve computational efficiency by exploiting specific properties of certain constraints and designing tailored algorithms accordingly. However, when a problem variant lacks the particular constraints, modifies the particular constraints, or introduces additional new constraints, the algorithmic structure designed for the original particular constraint may become incompatible.
A typical example is the Priority List (PL) method\cite{faizahDynamicPriorityApproach2024, elsayedThreestagePriorityList2021}, a widely used approach for UC. PL takes into account the load balance constraint and output power limits to quickly determine the on/off status of units. However, PL ignores startup/shutdown costs and ramping constraints. This omission not only leads to suboptimal solutions but also makes the resulting schedules impractical for real-world operation: disregarding ramping limits may require units to change output power too abruptly, making it physically impossible to follow the schedule. In the worst case, such aggressive dispatch instructions could even damage the units.
To fix this limitation, many subsequent extensions of the priority list method incorporate additional procedures\cite{faizahDynamicPriorityApproach2024, laoImplementationUnitCommitment2022, zhuTransferbasedApproximateDynamic2024}, such as economic dispatch or dynamic programming, to handle constraints originally neglected by the basic PL method. These extensions are often customized for specific problem variants.

The dependence of algorithmic parameters on problem parameters manifests itself in the need to adjust algorithmic parameters whenever problem parameters change, in order to obtain an optimal or even feasible solution. A typical example is the Lagrangian Relaxation (LR) method\cite{sunLagrangianDecompositionApproach2016}, a classical approach for solving the unit commitment problem. LR relaxes the power balance constraint, decomposing the original problem into independent single-unit subproblems, and then employs a subgradient method to iteratively update the Lagrangian multipliers in search of an optimal solution. However, its performance is not inherently guaranteed by the algorithm itself, but critically depends on the step-size strategy in the subgradient method, which must be manually tuned by the user for specific load or cost parameters. When the load fluctuates significantly, LR often struggles to find a feasible solution, necessitating the design of different step-size strategies for different load scenarios.

\section{Mathematical Models of UC and its Varaints}\label{sec:model}

The goal of the classical Unit Commitment Problem is to determine the on/off status and output power of each unit, minimize the total operating cost of the system while meeting load demand and system operational constraints. The mathematical model of this problem\cite{kerrUnitCommitment1966} is presented in Formulas~\eqref{eq:UC-1}-\eqref{eq:UC-7}.
\begin{align}
\min\quad & \sum_{i=1}^N \left[ \sum_{t=1}^T ( a_i S_{i,t} + b_i P_{i,t} + c_i P_{i,t}^2) \right. \notag \\
&\qquad \qquad \qquad \left. + \sum_{t=2}^T C_{i,t} (1-S_{i,t-1}) S_{i,t} \right] \label{eq:UC-1} \\
\text{s.t.}\quad & P_{i}^{min} S_{i,t} \le P_{i,t} \le P_{i}^{max} S_{i,t} \notag \\
&\qquad \qquad \qquad t =1,2,\dots,T,\quad i = 1,2,\dots,N \label{eq:UC-2} \\
& \sum_{i=1}^N S_{i,t} P_{i,t} = R_t \qquad \qquad \qquad t = 1,2,\dots,T \label{eq:UC-3} \\
& (X_{i,t-1}^{on} - H_{i}^{on})(S_{i,t-1} - S_{i,t}) \ge 0 \notag \\
&\qquad \qquad \qquad t = 2,3,\dots,T,\quad i = 1,2,\dots,N \label{eq:UC-4} \\
& (X_{i,t-1}^{off} - H_{i}^{off})(S_{i,t} - S_{i,t-1}) \ge 0 \notag \\
&\qquad \qquad \qquad t = 2,3,\dots,T,\quad i = 1,2,\dots,N \label{eq:UC-5} \\
& \sum_{i=1}^N S_{i,t} P_{i}^{max} \ge R_t E \qquad \qquad t = 1,2,\dots,T \label{eq:UC-6} \\
& C_{i,t} =
\begin{cases}
C_{i}^{cold}, & \operatorname{if} X_{i,t-1}^{off} \ge H_{i}^{off} + T_{i}^{cold} \\
C_{i}^{hot}, & \operatorname{otherwise}
\end{cases} \notag \\
&\qquad \qquad t = 2,3,\dots,T,\quad i = 1,2,\dots,N \label{eq:UC-7}
\end{align}
where $N$ and $T$ denotes the number of units and the number of scheduling time periods, respectively.
$ S_{i,t} $ is a binary variable indicating the on/off status of unit $ i $ at time $ t $.
$ P_{i,t} $ is the output power of unit $ i $ at time $ t $.
$ a_i $, $ b_i $, and $ c_i $ are cost coefficients for unit $ i $. $ C_{i,t} $ is the startup cost of unit $ i $ at time $ t $.
$ C_{i}^{cold} $ and $ C_{i}^{hot} $ represent the cold-start and hot-start costs of unit $ i $, respectively.
$ T_{i}^{cold} $ is the cold-start threshold (minimum offline time required for a cold start) of unit $ i $.
$ P_{i}^{min} $ and $ P_{i}^{max} $ are the minimum and maximum output power limits of unit $ i $, respectively. 
$ X_{i,t}^{on} $ and $ X_{i,t}^{off} $ represent the duration that unit $ i $ has been continuously online and offline, respectively, up to time $ t $.
$ H_{i}^{on} $ and $ H_{i}^{off} $ are the minimum up time and minimum down time of unit $ i $, respectively.
$ E $ is the reserve margin factor for load demand.
The objective function, denoted as Formula~\eqref{eq:UC-1}, minimizes the total operating cost of the power system.
Constraint~\eqref{eq:UC-2} enforces that the output power of each unit stays within its allowable limits.  
Constraint~\eqref{eq:UC-3} ensures that the total output power at each time meets the load demand.  
Constraints~\eqref{eq:UC-4}-\eqref{eq:UC-5} guarantee that the online and offline durations of each unit satisfy the minimum up time and minimum down time requirements. 
Constraint~\eqref{eq:UC-6} requires that the maximum available output power from running units at each time satisfies the spinning reserve requirement.  
Constraint~\eqref{eq:UC-7} specifies that the startup cost of a unit depends on its startup state.

The main difference between the classical UC\cite{kerrUnitCommitment1966} and its variants\cite{correa-posadaDynamicRampingModel2017, jinDataDrivenLookAheadUnit2019, liuAccurateModelingConfiguration2020, elsayedThreestagePriorityList2021, faizahDynamicPriorityApproach2024, karabasExactSolutionMethod2023, sunLagrangianDecompositionApproach2016, xavierTransmissionConstraintFiltering2019, ahmadiSecurityConstrainedUnitCommitment2019, gaoInternallyInducedBranchandcut2022, quLinearizationMethodLargescale2024, duHighefficiencyNetworkconstrainedClustered2019, wuAcceleratingNCUCBinary2016} lies in some additional constraints in the mathematical model of the variant, such as ramping constraints\cite{correa-posadaDynamicRampingModel2017, jinDataDrivenLookAheadUnit2019, liuAccurateModelingConfiguration2020, elsayedThreestagePriorityList2021, faizahDynamicPriorityApproach2024, karabasExactSolutionMethod2023}.
The UC with ramping constraints accounts for output power limitations under different operating conditions:
(1) When a unit is started up or shut down, its output power must equal to its minimum output power limitation.
(2) When a unit runs continuously, the change in its output power between consecutive time periods must not exceed the specified ramping rate.
Specifically, the mathematical model of the UC with ramping constraints\cite{correa-posadaDynamicRampingModel2017, jinDataDrivenLookAheadUnit2019, liuAccurateModelingConfiguration2020, elsayedThreestagePriorityList2021, faizahDynamicPriorityApproach2024, karabasExactSolutionMethod2023} introduces the additional constraints shown in Formulas~\eqref{eq:UCR-1}-\eqref{eq:UCR-4}.
\begin{align}
& y_{i,t} - z_{i,t} = S_{i,t}-S_{i,t-1} \notag \\
&\qquad \qquad t = 2,3,\dots,T,\quad i = 1,2,\dots,N \label{eq:UCR-1} \\
& y_{i,t} + z_{i,t} \le 1 \notag \\
&\qquad \qquad t = 2,3,\dots,T,\quad i = 1,2,\dots,N \label{eq:UCR-2} \\
& P_{i,t} - P_{i,t-1} \le S_{i,t-1} R_{i}^{up} + y_{i,t} P_{i}^{up} \notag \\
&\qquad \qquad t = 2,3,\dots,T,\quad i = 1,2,\dots,N \label{eq:UCR-3} \\
& P_{i,t-1} - P_{i,t} \le S_{i,t} R_{i}^{down} + z_{i,t} P_{i}^{down} \notag \\
&\qquad \qquad t = 2,3,\dots,T,\quad i = 1,2,\dots,N \label{eq:UCR-4}
\end{align}
where the newly introduced binary variables $ y_{i,t} $ and $ z_{i,t} $ indicate whether unit $ i $ starts up or shuts down at time $ t $, respectively. $ P_{i}^{up} $ and $ P_{i}^{down} $ denote the startup and shutdown ramping power of unit $ i $, with $ P_{i}^{up} = P_{i}^{down} = P_{i}^{min} $. $ R_{i}^{up} $ and $ R_{i}^{down} $ represent the maximum upward and downward ramping rates for unit $ i $. 
Constraint~\eqref{eq:UCR-1} ensures that $ y_{i,t} = 1 $ if unit $ i $ starts up at time $ t $, and $ z_{i,t} = 1 $ if it shuts down at time $ t $.  
Constraint~\eqref{eq:UCR-2} guarantees that $ y_{i,t} $ and $ z_{i,t} $ cannot both be equal to 1 simultaneously.  
Constraints~\eqref{eq:UCR-3}–\eqref{eq:UCR-4} enforce that the change in output power of unit $ i $ between consecutive time periods does not violate the upward and downward ramping rate limits.

The fundamental reason existing algorithms fail to generalize across diverse UC variants lies in their coupling with the specific mathematical model of each variant. The mathematical models of different UC variants inherently differ, particularly in the types and structures of operational constraints they include. As illustrated by the two UC formulations discussed above, the distinctions between the classical UC and its variants are directly encoded in their respective mathematical models as specific constraints. To accommodate these specific constraints, specialized algorithms often embed custom designed mechanisms tailored to specific constraints, thereby reinforcing the coupling between the algorithm and the problem.
However, we also note a key commonality: both problems can be formally expressed as mathematical models. This observation points to a promising pathway forward. By taking the mathematical model as the starting point, we can construct a unified solution framework capable of handling both the classical UC and its many variants.

\section{Unified Solution Framework for UC and its Variants}\label{sec:method}

To mitigate the generalizability limitation caused by the coupling of algorithms with specific problem variants in UC, this paper proposes a unified solution framework for UC and its variants as illustrated in Figure~\ref{fig:UniUC}.
Taking the mathematical model of UC as input, this framework automatically reduces any instance of the UC or its variants into a SAT instance, which is then solved by an SAT solver, thereby enabling a unified solution approach across different problem variants.
The underlying principle is that, although the mathematical models of different UC variants differ, every mathematical constraint in these models can be equivalently encoded as a set of SAT clauses. At this level of representation, there is no fundamental distinction exists between the classical UC and its variants. Furthermore, all resulting SAT instances conform to the standard SAT format. Consequently, there is no need to design specialized algorithms for individual variants. Instead, a standard SAT solver can be directly applied to uniformly solve both the classical UC and its many variants.
This framework effectively decouples the solution algorithm from the specific mathematical model of the problem.

\subsection{SAT-based Reduction for UC}

SAT\cite{chuanluoClauseStatesBased2015} is a classical problem in computer science and mathematical logic, first introduced by Stephen A. Cook\cite{cookComplexityTheoremprovingProcedures1971}. The goal of SAT is to determine whether there exists an assignment of truth values to Boolean variables that satisfies a given Boolean logical formula, i.e., makes the formula evaluate to true.  
Typically, the propositional formula is expressed in conjunctive normal form (CNF), where the formula is represented as a conjunction of clauses, and each clause is a disjunction of literals (a literal being either a Boolean variable or its negation). For example, the formula $(x_1 \lor \lnot x_2) \land (\lnot x_1 \lor x_2 \lor x_3)$ is a propositional formula in CNF. For notational convenience, we omit the conjunction symbols ``$\land$‘’ when describing CNFs in this paper, representing a CNF simply as a list of its clauses. For instance, the CNF $(x_1 \lor \lnot x_2) \land (\lnot x_1 \lor x_2 \lor x_3)$ will be denoted as Formulas~\eqref{eq:cnf-example-1}-\eqref{eq:cnf-example-2} in this paper.
\begin{gather}
x_1 \lor \lnot x_2 \label{eq:cnf-example-1} \\
\lnot x_1 \lor x_2 \lor x_3 \label{eq:cnf-example-2}
\end{gather}

SAT is chosen as the target for reducing UC because SAT is an NP-complete problem\cite{cookComplexityTheoremprovingProcedures1971}. This property implies that every problem in NP can be reduced to SAT in polynomial time. Notably, when the objective function of UC is disregarded, the task of verifying whether a given solution satisfies all its constraints belongs to NP. Consequently, the constraints of all variants of UC can, in theory, be reduced to a CNF instance of SAT. This provides a theoretical guarantee for a unified solution framework applicable to UC and its variants.

The SAT-based reduction pipeline for UC is illustrated in Figure~\ref{fig:UniUC}. The core procedure involves translating each constraint of UC into CNF and converting the objective function into an objective vector. This transformation maps diverse formulations of UC, despite their varying mathematical structures, into a standardized SAT representation. In this way, SAT-based reduction addresses the coupling between problem and algorithm that arises from structural differences among the mathematical models of UC and its variants.

In the SAT-based reduction, the reduction of constraint consists of two steps: (1) reducing constraints to the conjunction of propositional formulas, and (2) reducing the resulting conjunction into CNF. These two steps are described in detail below.

\subsection{Reduction to the Conjunction of Propositional Formulas}

In this section, we will take the capacity constraint of the classical UC, shown in Formula~\eqref{eq:UC-constraint-example-1}, as an example to illustrate how to translate problem constraints into the conjunction of propositional formulas from the perspective of the mathematical model.
\begin{align}
&P_{i}^{min} S_{i,t} \le P_{i,t} \le P_{i}^{max} S_{i,t} \notag \\
&\qquad \qquad t = 1,2,\dots,T,\quad i = 1,2,\dots,N \label{eq:UC-constraint-example-1}
\end{align}
where $N$ and $T$ denotes the number of units and the number of scheduling time periods, respectively.
$ S_{i,t} $ is a binary variable indicating the on/off status of unit $ i $ at time $ t $.
$ P_{i,t} $ is the output power of unit $ i $ at time $ t $.
$ P_{i}^{min} $ and $ P_{i}^{max} $ are the minimum and maximum output power limits of unit $ i $, respectively. 

Automated reduction methods\cite{janicicURSASystemUniform2012, semenovTranslationAlgorithmicDescriptions2020} typically introduce auxiliary variables $ \mathcal{P}^{min}_{i,t} $ and $ \mathcal{P}^{max}_{i,t} $ to represent the values of $ P_{i}^{min} S_{i,t} $ and $ P_{i}^{max} S_{i,t} $, respectively. Consequently, the constraint~\eqref{eq:UC-constraint-example-1} can be transformed into the conjunction of propositional formulas as shown in Formulas~\eqref{eq:UC-constraint-origin-rule-1}-\eqref{eq:UC-constraint-origin-rule-4}.

\begin{gather}
\bigwedge_{t=1}^T \bigwedge_{i=1}^N \left( P_{i}^{min} S_{i,t} = \mathcal{P}_{i,t}^{min} \right) \label{eq:UC-constraint-origin-rule-1} \\
\bigwedge_{t=1}^T \bigwedge_{i=1}^N \left( P_{i}^{max} S_{i,t} = \mathcal{P}_{i,t}^{max} \right) \label{eq:UC-constraint-origin-rule-2} \\
\bigwedge_{t=1}^T \bigwedge_{i=1}^N \left( \mathcal{P}_{i,t}^{min} \le P_{i,t} \right) \label{eq:UC-constraint-origin-rule-3} \\
\bigwedge_{t=1}^T \bigwedge_{i=1}^N \left( P_{i,t} \le \mathcal{P}_{i,t}^{max} \right) \label{eq:UC-constraint-origin-rule-4}
\end{gather}

This reduction rule enables a relatively simple and generic reduction of constraints into propositional formulas that are close to CNF. However, this rule introduces a large number of auxiliary variables. Besides, the multiplication operations involved, when reduced to SAT, will generate an excessive number of Boolean variables and clauses, resulting in an overly large SAT instance that hinders efficient solving.

To mitigate this issue, the reduction rule can be adapted based on the structural properties of the mathematical model. For instance, in the case of the power output constraint discussed above, Formula~\eqref{eq:UC-constraint-example-1} reveals that when a unit is offline ($S_{i,t} = 0$), its output power $P_{i,t}$ must be zero; only when the unit is online ($S_{i,t} = 1$), $P_{i,t}$ is bounded by the minimum and maximum power limits $P_{i}^{min}$ and $P_{i}^{max}$. Accordingly, the constraint~\eqref{eq:UC-constraint-example-1} can be reformulated as the conjunction of propositional formulas shown in Formulas~\eqref{eq:UC-constraint-final-rule-1}-\eqref{eq:UC-constraint-final-rule-2}.
\begin{gather}
\bigwedge_{t=1}^T \bigwedge_{i=1}^N \left( (S_{i,t} = 0) \to (P_{i,t}=0) \right) \label{eq:UC-constraint-final-rule-1} \\
\bigwedge_{t=1}^T \bigwedge_{i=1}^N \left( (S_{i,t} = 1) \to \left( (P_{i}^{min} \le P_{i,t}) \land (P_{i,t} \le P_{i}^{max}) \right) \right) \label{eq:UC-constraint-final-rule-2}
\end{gather}
Let $\alpha$, $\beta$ and $\gamma$ denote three propositions, such as ``$S_{i,t}=0$''. Since $\alpha \to \beta$ is equivalent to $\neg \alpha \lor \beta$, Formula~\eqref{eq:UC-constraint-final-rule-1} can be transformed into Formula~\eqref{eq:UC-constraint-final-rule-3}. Additionally, since $\alpha \to (\beta \land \gamma)$ is logically equivalent to the conjunction of $\alpha \to \beta$ and $\alpha \to \gamma$, Formula~\eqref{eq:UC-constraint-final-rule-2} can be transformed into Formulas~\eqref{eq:UC-constraint-final-rule-4}-\eqref{eq:UC-constraint-final-rule-5}.

\begin{gather}
\bigwedge_{t=1}^T \bigwedge_{i=1}^N \left( (S_{i,t} = 1) \lor (P_{i,t}=0) \right) \label{eq:UC-constraint-final-rule-3} \\
\bigwedge_{t=1}^T \bigwedge_{i=1}^N \left( (S_{i,t} = 0) \lor (P_{i}^{min} \le P_{i,t}) \right) \label{eq:UC-constraint-final-rule-4} \\
\bigwedge_{t=1}^T \bigwedge_{i=1}^N \left( (S_{i,t} = 0) \lor (P_{i,t} \le P_{i}^{max}) \right) \label{eq:UC-constraint-final-rule-5}
\end{gather}

As shown in Formulas~\eqref{eq:UC-constraint-final-rule-1}-\eqref{eq:UC-constraint-final-rule-5}, by tailoring the reduction process to the structure of the mathematical model, we eliminate explicit auxiliary variables and multiplication operations. This reduction rule further reduce the number of propositional formulas, yielding a smaller-scale SAT instance that is more easy to solving.

\subsection{Reduction to CNF}

The conjunctions of propositional formulas, such as Formulas~\eqref{eq:UC-constraint-final-rule-3}-\eqref{eq:UC-constraint-final-rule-5}, are already very close to CNF. However, the propositional formulas still contain elements not permitted in standard CNF such as numerical variables, arithmetic operations, and numerical comparisons. To convert these formulas into a CNF-compatible representation, each numerical variable is encoded as a Boolean vector corresponding to its fixed-point binary representation, as illustrated in Formula~\eqref{eq:variable-encoding}.
\begin{gather}
(p_{1}, \cdots, p_{n},p_{n+1},\cdots,p_{n+m}) \label{eq:variable-encoding}
\end{gather}
where $ n $ denotes the number of integer bits and $ m $ the number of fractional bits. The bits $ p_1, \dots, p_n $ represent the integer part of the encoded value, while the bits $ p_{n+1}, \dots, p_{n+m} $ represent its fractional part. The numerical value encoded by this Boolean vector is $\sum_{i=1}^{n+m} 2^{\,n-i} \, p_i$. 
Since all variables in UC are non-negative, this fixed-point binary encoding is sufficient for the reduction. Consequently, numerical variables in Formula~\eqref{eq:UC-constraint-final-rule-3}-\eqref{eq:UC-constraint-final-rule-5}, such as $ P_{i}^{min} $, $ P_{i}^{max} $, and $ P_{i,t} $, can all be represented as Boolean vectors. In contrast, Boolean variables such as $ S_{i,t} $ require no additional encoding and remain as native Boolean variables.

For arithmetic operations, we encode addition and subtraction using full-adder circuits\cite{zhouOptimizingSATEncodings2017}, while encoding multiplication and division using multiplier circuits, thereby supporting basic computational requirements.

For numerical comparisons, a common approach is to apply Tseitin transformation\cite{janicicURSASystemUniform2012} to obtain CNF encodings. Taking $ X > Y $ as an example, let the fixed-point Boolean encodings of the numerical variables $ X $ and $ Y $ be $ (x_1, \dots, x_{n+m}) $ and $ (y_1, \dots, y_{n+m}) $, respectively. A Tseitin-based reduction first reduce the comparison $ X > Y $ as Formula~\eqref{eq:compare-tseitin-1}.
\begin{gather}
(x_1 \land \neg y_1) \lor \bigvee_{i=2}^{n+m} \left( (x_i \land \neg y_i) \land \bigwedge_{j=1}^{i-1} (x_j \leftrightarrow y_j) \right) \label{eq:compare-tseitin-1}
\end{gather}
The above expression is then decomposed into CNF using the Tseitin transformation. As an example, Formulas~\eqref{eq:compare-tseitin-start}-\eqref{eq:compare-tseitin-end} consider the case where $n + m = 2$ :
\begin{gather}
u_1 \label{eq:compare-tseitin-start} \\
u_1 \leftrightarrow (u_2 \lor u_3) \label{eq:compare-tseitin-second} \\
u_2 \leftrightarrow (x_1 \land \neg y_1) \\
u_3 \leftrightarrow (u_4 \land u_5) \\
u_4 \leftrightarrow (u_6 \lor u_7) \\
u_5 \leftrightarrow (x_2 \land \neg y_2) \\
u_6 \leftrightarrow (x_1 \land y_1) \\
u_7 \leftrightarrow (\neg x_1 \land \neg y_1) \label{eq:compare-tseitin-end}
\end{gather}
where $ u_i $ denotes an auxiliary variable introduced by the Tseitin transformation. Let $a$, $b$ and $c$ be three literals in Formulas~\eqref{eq:compare-tseitin-second}-\eqref{eq:compare-tseitin-end}, such as $u_2$, $x_1$ and $\neg y_1$. The formula $a \leftrightarrow (b \lor c)$ is equivalent to the following three clauses in CNF: $\neg a \lor b \lor c$, $\neg b \lor a$ and $\neg c \lor a$. Similarly, the formula $a \leftrightarrow (b \land c)$ is equivalent to three CNF clauses: $\neg a \lor b$, $\neg a \lor c$ and $\neg b \lor \neg c \lor a$. Following these two transformation rules, Formulas~\eqref{eq:compare-tseitin-second}-\eqref{eq:compare-tseitin-end} can be fully converted into CNF. For instance, Formula~\eqref{eq:compare-tseitin-second} can be transformed into three CNF clauses: $\neg u_1 \lor u_2 \lor u_3$, $\neg u_2 \lor u_1$ and $\neg u_3 \lor u_1$.

Tseitin-based reduction can automatically convert any propositional formula into CNF. However, we observe that the resulting SAT instances are often inefficient to solve.  
Take the CNF encoding of $ X > Y $, shown in Formulas~\eqref{eq:compare-tseitin-start}-\eqref{eq:compare-tseitin-end}, as an example. If the most significant bits satisfy $ x_1 > y_1 $, i.e., $ x_1 = 1 $ and $ y_1 = 0 $, it is immediately clear that $ X > Y $ holds. When these values are substituted into Formulas~\eqref{eq:compare-tseitin-start}-\eqref{eq:compare-tseitin-end}, a SAT solver can quickly infer via unit propagation that $ u_1 = u_2 = 1 $ and $ u_3 = u_4 = u_6 = u_7 = 0 $. However, the clauses encoding $ u_5 \leftrightarrow (x_2 \land \neg y_2) $ remain active, even though they are irrelevant to establishing $ X > Y $ under the given assignment.  
The solver must still spend time making branching decisions on $ u_5 $, $ x_2 $, or $ y_2 $ to satisfy this redundant clause, which degrades solving efficiency.  
This observation suggests a key principle: the CNF generated by reduction should enable unit propagation to rapidly eliminate clauses that are no longer relevant to the current search context, thereby avoiding the negative impact of such inactive clauses on solver performance.

To address the low solving efficiency of SAT instances generated by the Tseitin transformation for numerical comparisons, we propose a binary-comparison-based reduction rule. Taking $ X > Y $ as an example, our rule produces the CNF shown in Formulas~\eqref{eq:compare-opt-1}-\eqref{eq:compare-opt-15}.
\begin{align}
(x_1 \land \neg y_1) \to T_1 \label{eq:compare-opt-1} \\
(x_1 \land y_1) \to E_1 \label{eq:compare-opt-2} \\
(\neg x_1 \land \neg y_1) \to  E_1 \label{eq:compare-opt-3} \\
(\neg x_1 \land y_1) \to \neg E_1 \label{eq:compare-opt-4} \\
(\neg x_1 \land y_1) \to \neg T_1 \label{eq:compare-opt-5} \\
\neg(T_1 \land E_1) \label{eq:compare-opt-6} \\
\neg E_{i-1} \to \neg E_i, && i = 2,3,\dots,n+m \label{eq:compare-opt-7} \\
\neg E_{i-1} \to \neg T_i, && i = 2,3,\dots,n+m \label{eq:compare-opt-8} \\
(E_{i-1} \land x_i \land \neg y_i) \to T_i, && i = 2,3,\dots,n+m \label{eq:compare-opt-9} \\
(E_{i-1} \land x_i \land y_i) \to E_i, && i = 2,3,\dots,n+m \label{eq:compare-opt-10} \\
(E_{i-1} \land \neg x_i \land \neg y_i) \to E_i, && i = 2,3,\dots,n+m \label{eq:compare-opt-11} \\
(E_{i-1} \land \neg x_i \land y_i) \to \neg E_i, && i = 2,3,\dots,n+m \label{eq:compare-opt-12} \\
(E_{i-1} \land \neg x_i \land y_i) \to \neg T_i, && i = 2,3,\dots,n+m \label{eq:compare-opt-13} \\
\neg(E_{i-1} \land T_i \land E_i), && i = 2,3,\dots,n+m \label{eq:compare-opt-14} \\
\bigvee_{i=1}^{n+m} T_i \label{eq:compare-opt-15} 
\end{align}
where $ T_i $ and $ E_i $ are auxiliary variables. For $ i = 1 $, $ T_1 $ denotes $ x_1 \land \neg y_1 $, $ E_1 $ denotes $ x_1 \leftrightarrow y_1 $. For $ i \ge 2 $, $ T_i $ denotes $ (x_i \land \neg y_i) \land \bigwedge_{j=1}^{i-1} (x_j \leftrightarrow y_j) $, $ E_i $ denotes $ \bigwedge_{j=1}^{i} (x_j \leftrightarrow y_j) $.
Once the values of $ x_1 $ and $ y_1 $ are determined, $ T_1 $ and $ E_1 $ can be immediately inferred via unit propagation. When propagating from the most significant bit downward, if at position $ i $ we encounter $ x_i \neq y_i $, then $ E_i $ is forced to 0. According to Formulas~\eqref{eq:compare-opt-7}-\eqref{eq:compare-opt-8}, this implies that for all $ j > i $, both $ E_j = 0 $ and $ T_j = 0 $. As a result, all remaining clauses in the CNF encoding of $ X > Y $ are satisfied through unit propagation, requiring no additional branching decisions. This avoids wasted computational effort by the SAT solver on redundant clauses that no longer affect the truth value of the comparison.

Following the above reduction process, all constraints in the mathematical model are transformed into CNF and incorporated into the final SAT instance. As for the objective function, we can introduce an auxiliary variable $ O $ and equate $O$ to the objective expression, thereby treating the equation as an additional constraint. The computation of $ O $ can then be encoded into CNF using the same reduction rule as constraints, yielding a Boolean vector $ (o_1, \dots, o_{n+m}) $ that represents the encoded objective value $O$.

\subsection{Algorithm for Optimal SAT Solution}

Through SAT-based reduction, we obtain a SAT instance $\Psi$. However, this SAT instance can only yield feasible solutions. To find an optimal solution, we require an additional optimization algorithm that minimizes the objective value $ O $ while ensuring CNF $ \Psi $ remains satisfied.

Thanks to the SAT-based reduction, the input to the optimization algorithm, across all variants of the unit commitment problem, is uniformly structured: a standardized SAT formula $ \Psi $ together with an objective variable $ O $. Consequently, the algorithm design no longer needs to account for the diverse constraints of the original unit commitment formulations. Instead, a single algorithm tailored to this unified SAT representation suffices to solve all problem variants, thereby eliminating coupling between the optimization algorithm and the original problem structure.

\begin{algorithm}[!t]
\footnotesize
\caption{Algorithm for Optimal SAT Solution}
\label{alg:sat-optimal-solution-search-algorithm}
\begin{algorithmic}[1]
\Require $\Psi$: the reduced SAT instance, $O$: the objective variable encoded in $\Psi$
\Ensure $V_{latest}$: the optimal solution
\State $V_{latest} = \emptyset$
\State solve $\Psi$ with a standard SAT solver, get the output assignment $V$
\While{$V \neq \emptyset$} \Comment{not UNSAT}
  \State $V_{latest} = V$
  \State $\Omega =$ the value of $O$ in assigment $V$ \Comment{the latest founded minimal objective value}
  \State reduce ``$O < \Omega$'' as a contraint into CNF as $\psi$ \Comment{using SAT-based reduction}
  \State $\Psi = \Psi \land \psi$
  \State solve $\Psi$ with a standard SAT solver, get the output assignment $V$
\EndWhile
\State \Return $V_{latest}$
\end{algorithmic}
\end{algorithm}

We employ a linear-search-based algorithm as the optimization algorithm\cite{aloulSearchTechniquesSATbased2006}, as shown in Algorithm~\ref{alg:sat-optimal-solution-search-algorithm}.  
First, we solve the reduced SAT instance $\Psi$ using a standard SAT solver to obtain an assignment $V$, i.e., a feasible solution to $\Psi$. If $V = \emptyset$, it means the solver fails to find any feasible solution.  
Whenever the solver returns a feasible solution $V$, we compute its corresponding objective value $\Omega$. We then add a new constraint $O < \Omega$, which is subsequently reduced to CNF and incorporated into $\Psi$. This additional constraint tightens the feasible region by excluding solutions with objective values no better than $\Omega$, while ensuring that all remaining feasible solutions have strictly improved objective values.  
We repeat this process, solving the updated $\Psi$, extracting $\Omega$, and adding the new constraint $O < \Omega$, until the SAT solver reports no further feasible solution. At that point, the last feasible solution found is guaranteed to be optimal.

\section{Experiments}\label{sec:experiments}

In this section, we evaluate the generalizability of the proposed unified solution framework.
Experiments on the classical UC and its variant with ramping constraints show that the proposed method outperforms existing specialized algorithms on both problems. The result demonstrates stronger generalizability across problem variants.
With this framework, new UC variants no longer require a custom algorithm design. These variants can be solved by simply adjusting the input model, improving the adaptability and response speed of power systems in diverse operational scenarios.

\subsection{Benchmark}

In this paper, we construct two benchmarks: UNIT\_N\_T for the classical UC and UNIT\_N\_T\_RAMP for the UC with ramping constraints. Each benchmark consists of publicly available benchmark instances collected from the literature\cite{faizahDynamicPriorityApproach2024, aliSolvingFuelBasedUnit2024, vatambetiSynergisticOptimizationUnit2024, castilloUnitCommitmentProblem2016, kambojNovelHybridPSO2016} and additional synthetic instances generated to supplement the collection. The synthetic instances are produced using the open-source, realistic UC instance generator\cite{borghettiLagrangianHeuristicsBased2002} available at \url{https://commalab.di.unipi.it/datasets/UC/}.
The UNIT\_N\_T benchmark contains 27 instances. The number of units ranges from 2 to 10, increasing incrementally by one unit per group. For each unit size, three different scheduling horizons, 8, 16, and 24 time periods, are considered, resulting in three variants per unit size.

The UNIT\_N\_T\_RAMP benchmark follows the exact same structure as UNIT\_N\_T, with the only difference being the inclusion of ramping parameters. These ramping limits are added according to the methodology used in the public benchmarks: for each unit, ramp rates are randomly generated in proportion to its capacity, scaled by a margin factor uniformly sampled from the interval $[0.9, 1.1]$.
All benchmark instances is available at \url{https://anonymous.4open.science/r/UniUC/}.

\subsection{Experimental Settings}
The experiments in this paper were conducted on a machine equipped with an Intel(R) Xeon(R) Emerald Rapids CPU and 32 GB of RAM. We employed the SAT solver CryptoMiniSat\cite{heuleSATCompetition2018}, leveraging its incremental solving capability to enhance computational efficiency. A time limit of 8 hours was imposed for each run. If the algorithm did not terminate within this limit, only the best feasible solution found up to that point was returned.

\subsection{Generalizability}

To evaluate the generalizability of the proposed unified solution framework for the UC and its variants, we conduct experimental comparisons against existing specialized algorithms designed for the classical UC and its variant with ramping constraints.
For the classical UC, we select three representative algorithms specifically designed or adapted for this problem: the Bald Eagle Search (BES) algorithm\cite{aliSolvingFuelBasedUnit2024}, the Branch-and-Bound Method (BBM)\cite{ZhangBranchBoundMethod2025}, and the Particle Swarm Optimization (PSO)\cite{vatambetiSynergisticOptimizationUnit2024}.
For the UC with ramping constraints, we choose three dedicated algorithms commonly used in power systems: the Dynamic Priority Approach with Particle Swarm Optimization (DPA-PSO)\cite{faizahDynamicPriorityApproach2024}, the Priority List (PL) method\cite{elsayedThreestagePriorityList2021}, and the Lagrangian Relaxation (LR) algorithm\cite{saranyaSolutionUnitCommitment2019}.
We perform experiments on 27 publicly available test instances for each problem. The benchmarks for the classical UC and the UC with ramping constraints are denoted as UNIT\_N\_T and UNIT\_N\_T\_RAMP, respectively.
To mitigate random variation, each algorithm is executed 30 times on every instance. For each algorithm-instance pair, we record the average objective value (AVG) and the best objective value (BEST) across the 30 runs. Based on these results, we compute the average rank of each algorithm under both the AVG and BEST metrics, as summarized in Tables~\ref{tab:comparision-in-unit_N_T}-\ref{tab:comparision-in-unit_N_T_R}.

{
\begin{table*}[p]
\centering
\caption{Comparison on UNIT\_N\_T for the classical UC}
\label{tab:comparision-in-unit_N_T}
\begin{tabular}{c r r r r r r r r}
\toprule
Instance & \multicolumn{2}{c}{BES} & \multicolumn{2}{c}{BBM} & \multicolumn{2}{c}{PSO} & \multicolumn{2}{c}{OURS} \\
\cmidrule(lr){2-3} \cmidrule(lr){4-5} \cmidrule(lr){6-7} \cmidrule(lr){8-9} 
& \multicolumn{1}{c}{AVG} & \multicolumn{1}{c}{BEST} & \multicolumn{1}{c}{AVG} & \multicolumn{1}{c}{BEST} & \multicolumn{1}{c}{AVG} & \multicolumn{1}{c}{BEST} & \multicolumn{1}{c}{AVG} & \multicolumn{1}{c}{BEST} \\
\midrule
unit\_2\_8 & 22575.47 & 22575.47 & 21429.43 & 21429.43 & 22295.18 & 22295.18 & \textbf{21071.90} & \textbf{21071.90} \\
unit\_2\_16 & 46306.73 & 46306.73 & 47250.80 & 47105.80 & 48067.74 & 48067.74 & \textbf{44962.80} & \textbf{44962.80} \\
unit\_2\_24 & 72287.85 & 72287.85 & 71135.13 & 71135.13 & 73037.22 & 73037.22 & \textbf{70775.10} & \textbf{70775.10} \\
unit\_3\_8 & 22185.87 & 22185.87 & 18795.00 & 16963.77 & 22012.88 & 22012.88 & \textbf{16894.87} & \textbf{16894.87} \\
unit\_3\_16 & 52408.32 & 52408.32 & 50401.79 & 50401.79 & 54918.07 & 54669.78 & \textbf{50151.74} & \textbf{50151.74} \\
unit\_3\_24 & 83455.00 & 83402.00 & 82343.68 & 82343.68 & 88569.02 & 88184.85 & \textbf{80605.27} & \textbf{80605.27} \\
unit\_4\_8 & \textbf{73933.00} & \textbf{73933.00} & 74380.35 & 74380.35 & 74573.86 & 74431.86 & 74355.00 & 74216.00 \\
unit\_4\_16 & 21659.90 & 21659.90 & 30191.30 & 25912.14 & 21175.71 & 21175.71 & \textbf{20210.04} & \textbf{20210.04} \\
unit\_4\_24 & 14036.25 & 14036.25 & 23137.05 & 15161.18 & 15247.54 & 15247.54 & \textbf{14035.91} & \textbf{14035.91} \\
unit\_5\_8 & 4695.43 & 4678.27 & 4617.00 & 4617.00 & 4846.68 & 4713.86 & \textbf{4423.00} & \textbf{4423.00} \\
unit\_5\_16 & 4665.00 & 4661.00 & 4044.00 & 4013.00 & 5119.72 & 5119.72 & \textbf{3830.00} & \textbf{3830.00} \\
unit\_5\_24 & 12457.95 & 12413.09 & 25897.40 & 25897.40 & 14888.64 & 14716.39 & \textbf{12308.00} & \textbf{12272.00} \\
unit\_6\_8 & 5282.10 & 5274.91 & 4961.37 & 4961.37 & 5722.87 & 5610.01 & \textbf{4927.33} & \textbf{4927.33} \\
unit\_6\_16 & 17736.53 & 17721.79 & 18426.53 & 18426.53 & 25100.78 & 23930.20 & \textbf{17569.94} & \textbf{17569.94} \\
unit\_6\_24 & 15725.88 & 15687.15 & \textbf{13430.92} & \textbf{12866.75} & 14598.00 & 14598.00 & 13951.00 & 13823.00 \\
unit\_7\_8 & 13314.95 & 13314.56 & 11338.77 & 11296.77 & 13710.44 & 13710.44 & \textbf{9611.99} & \textbf{9611.99} \\
unit\_7\_16 & 24259.82 & 24259.82 & 38514.32 & 30516.11 & 24212.54 & 24212.54 & \textbf{23850.00} & \textbf{23850.00} \\
unit\_7\_24 & 40324.23 & 40321.39 & 41983.36 & 39179.36 & 40915.63 & 40915.63 & \textbf{32501.06} & \textbf{32501.06} \\
unit\_8\_8 & 5317.24 & 5274.91 & 5938.97 & 5938.97 & 6252.01 & 6006.60 & \textbf{5027.00} & \textbf{5027.00} \\
unit\_8\_16 & 25463.44 & 25436.72 & \textbf{24790.58} & \textbf{24790.58} & 41380.30 & 39907.62 & 25431.00 & 25431.00 \\
unit\_8\_24 & 42398.81 & 42193.85 & 39460.94 & 39460.94 & 70136.23 & 68347.74 & \textbf{37167.78} & \textbf{37167.78} \\
unit\_9\_8 & 4722.09 & 4678.21 & 5298.74 & 5298.74 & 5860.73 & 5536.25 & \textbf{4370.00} & \textbf{4370.00} \\
unit\_9\_16 & 25923.06 & 25886.68 & 25591.07 & 25591.07 & 46185.94 & 39617.80 & \textbf{25419.07} & \textbf{25419.07} \\
unit\_9\_24 & 36392.68 & 36260.27 & 35109.33 & 35109.00 & 77978.13 & 73958.42 & \textbf{34231.33} & \textbf{34231.33} \\
unit\_10\_8 & 10063.61 & 10063.61 & 10470.86 & 10470.86 & 13594.15 & 12659.90 & \textbf{9919.34} & \textbf{9919.34} \\
unit\_10\_16 & 15861.86 & 15853.86 & 16977.44 & 16977.44 & 36359.66 & 34754.35 & \textbf{15641.92} & \textbf{15641.92} \\
unit\_10\_24 & 563937.00 & 563937.00 & 563977.00 & 563977.00 & \textbf{563783.00} & \textbf{561284.00} & 583264.00 & 580475.00 \\
\midrule
Average Rank & \multicolumn{1}{c}{2.67} & \multicolumn{1}{c}{2.70} & \multicolumn{1}{c}{2.59} & \multicolumn{1}{c}{2.48} & \multicolumn{1}{c}{3.52} & \multicolumn{1}{c}{3.59} & \multicolumn{1}{c}{\textbf{1.22}} & \multicolumn{1}{c}{\textbf{1.22}} \\
\bottomrule
\end{tabular}
\end{table*}
}

{
\begin{table*}[p]
\centering
\caption{Comparison on UNIT\_N\_T\_RAMP for the UC with ramping constraints}
\label{tab:comparision-in-unit_N_T_R}
\begin{tabular}{c r r r r r r r r}
\toprule
Instance & \multicolumn{2}{c}{DPA-PSO} & \multicolumn{2}{c}{LR} & \multicolumn{2}{c}{PL} & \multicolumn{2}{c}{OURS} \\
\cmidrule(lr){2-3} \cmidrule(lr){4-5} \cmidrule(lr){6-7} \cmidrule(lr){8-9} 
& \multicolumn{1}{c}{AVG} & \multicolumn{1}{c}{BEST} & \multicolumn{1}{c}{AVG} & \multicolumn{1}{c}{BEST} & \multicolumn{1}{c}{AVG} & \multicolumn{1}{c}{BEST} & \multicolumn{1}{c}{AVG} & \multicolumn{1}{c}{BEST} \\
\midrule
unit\_2\_8\_R & 21412.05 & 21412.05 & 21897.76 & 21784.46 & 22415.23 & 22415.23 & \textbf{21071.90} & \textbf{21071.90} \\
unit\_2\_16\_R & 46306.71 & 46306.71 & 46203.48 & 46203.48 & 48067.74 & 48067.74 & \textbf{46076.89} & \textbf{46076.89} \\
unit\_2\_24\_R & 71151.57 & 71151.57 & 72753.93 & 72512.53 & 72891.33 & 72891.33 & \textbf{70775.10} & \textbf{70775.10} \\
unit\_3\_8\_R & \textbf{20438.67} & \textbf{20438.67} & 21799.77 & 21799.77 & 21719.78 & 21719.78 & \textbf{20438.67} & \textbf{20438.67} \\
unit\_3\_16\_R & 55529.48 & 55529.48 & 53256.08 & 52236.89 & 62349.56 & 62349.56 & \textbf{50151.74} & \textbf{50151.74} \\
unit\_3\_24\_R & 86872.65 & 86863.64 & 83730.90 & 82736.35 & 98253.85 & 98253.85 & \textbf{80605.27} & \textbf{80605.27} \\
unit\_4\_8\_R & 78246.32 & 78204.57 & 76444.00 & 75312.40 & 74812.00 & 74812.00 & \textbf{74355.00} & \textbf{74216.00} \\
unit\_4\_16\_R & 37376.55 & 37350.39 & 38500.00 & 38500.00 & 37997.28 & 37837.66 & \textbf{37337.00} & \textbf{37337.00} \\
unit\_4\_24\_R & 25225.33 & 25225.33 & 24979.47 & 24979.47 & 29445.92 & 28528.92 & \textbf{24830.00} & \textbf{24830.00} \\
unit\_5\_8\_R & 15444.92 & 15444.38 & 14616.96 & 12833.04 & 15726.72 & 15726.72 & \textbf{12810.42} & \textbf{12810.42} \\
unit\_5\_16\_R & 18851.88 & 18826.71 & 19338.26 & 19338.26 & 23033.47 & 22334.39 & \textbf{18433.85} & \textbf{18433.85} \\
unit\_5\_24\_R & 18486.29 & 18459.35 & \textbf{13281.53} & \textbf{12788.15} & 13475.00 & 13475.00 & 13891.00 & 13877.00 \\
unit\_6\_8\_R & 6273.63 & 6271.45 & 5408.00 & 5133.74 & 6325.54 & 6325.54 & \textbf{4927.36} & \textbf{4927.36} \\
unit\_6\_16\_R & 29197.50 & 29197.17 & 27600.94 & 17874.89 & 29879.71 & 29879.71 & \textbf{17569.94} & \textbf{17569.94} \\
unit\_6\_24\_R & \textbf{13560.27} & \textbf{13560.27} & 15506.11 & 15506.11 & 16277.79 & 16271.62 & 13951.00 & 13823.00 \\
unit\_7\_8\_R & 10070.20 & 10070.20 & 10470.86 & 10470.86 & 11862.32 & 11352.07 & \textbf{9918.98} & \textbf{9918.98} \\
unit\_7\_16\_R & 25280.53 & 25280.53 & 25054.30 & 25054.30 & 38834.27 & 37405.99 & \textbf{24727.96} & \textbf{24727.96} \\
unit\_7\_24\_R & 34065.53 & 34053.83 & 35799.62 & 35799.62 & 61982.35 & 59798.37 & \textbf{32642.81} & \textbf{32642.81} \\
unit\_8\_8\_R & 6254.60 & 6248.59 & 7947.40 & 7543.99 & 8496.53 & 8496.53 & \textbf{5027.00} & \textbf{5027.00} \\
unit\_8\_16\_R & 27323.49 & 27320.27 & 37304.50 & 34253.50 & 27491.17 & 27491.17 & \textbf{25431.81} & \textbf{25431.81} \\
unit\_8\_24\_R & 40946.88 & 40943.54 & 38228.03 & 37947.64 & 37885.15 & 37885.15 & \textbf{37167.78} & \textbf{37167.78} \\
unit\_9\_8\_R & 7183.22 & 7166.73 & 5037.70 & 4605.99 & 7432.70 & 7432.70 & \textbf{4370.00} & \textbf{4370.00} \\
unit\_9\_16\_R & 26241.73 & 26239.45 & 27429.00 & 26256.41 & 26440.29 & 26440.29 & \textbf{25572.00} & \textbf{25572.00} \\
unit\_9\_24\_R & 41251.21 & 41246.39 & 37031.15 & 36715.55 & 39311.36 & 39311.36 & \textbf{34231.33} & \textbf{34231.33} \\
unit\_10\_8\_R & 24795.94 & 24795.94 & 25515.02 & 23784.63 & 26611.57 & 26611.57 & \textbf{23064.03} & \textbf{23064.03} \\
unit\_10\_16\_R & 31770.70 & 31769.81 & 30993.50 & 27218.32 & 33681.33 & 33681.33 & \textbf{26214.05} & \textbf{26214.05} \\
unit\_10\_24\_R & 629198.83 & 627775.68 & 568071.30 & \textbf{563215.31} & \textbf{564534.00} & 564534.00 & 584931.00 & 581033.00 \\
\midrule
Average Rank & \multicolumn{1}{c}{2.69} & \multicolumn{1}{c}{2.72} & \multicolumn{1}{c}{2.63} & \multicolumn{1}{c}{2.52} & \multicolumn{1}{c}{3.48} & \multicolumn{1}{c}{3.56} & \multicolumn{1}{c}{\textbf{1.20}} & \multicolumn{1}{c}{\textbf{1.20}} \\
\bottomrule
\end{tabular}
\end{table*}
}

For the classical UC, Table~\ref{tab:comparision-in-unit_N_T} shows that the proposed unified solution framework achieves the best average rank compared to BES, BBM, and PSO.
On both the AVG and BEST metrics, the proposed framework yields higher solution quality than BES on 25 instances (25/27), outperforms BBM on 24 instances (24/27), and surpasses PSO on 26 instances (26/27).
Both BES and PSO incorporate operational characteristics of units into their design, employing tailored population update strategies. Such heuristic rules often yield strong performance on test instances that closely align with the underlying assumptions of heuristic rules \cite{huangCorrelationbasedDynamicAllocation2024, yangLocaldiversityEvaluationAssignment2023}. However, their generalizability tends to degrade when applied to instances that deviate significantly from these assumptions.
BBM addresses the nonlinearity and non-convexity of the quadratic objective function in UC by linearizing the objective and reformulating the model as a linear model, which is then solved using the high-performance branch-and-bound solver CPLEX\cite{IBMILOGCPLEXOptimizationStudio2024}. However, this linearization inevitably introduces approximation errors, causing the optimal solution of the linearized model to deviate from the optimum of the original problem. This error becomes more pronounced when the quadratic cost coefficients are large or the output power ranges are wide, leading the optimization process astray under an inaccurate objective landscape. Moreover, the heuristic strategy in BBM based on the heat rate of the unit may further bias the search away from the global optimum.
In contrast, the proposed unified solution framework relies on no domain-specific prior assumptions. Instead, it uniformly reduces all problem instances to SAT instances, thereby achieving consistently stable solution quality across diverse test instances.

For the UC with ramping constraints, Table~\ref{tab:comparision-in-unit_N_T_R} shows that the proposed unified solution framework achieves the best average rank among DPA-PSO, LR, and PL. On both the AVG and BEST metrics, the proposed framework achieves better solutions compared to DPA-PSO on 25 test instances (25/27), outperforms LR on 25 instances (25/27), and surpasses PL on 25 instances (25/27).
DPA-PSO and PL each construct distinct priority models to compute priority rankings, which are used to determine unit scheduling order. Such heuristic rules can accelerate the search process by prioritizing high-ranking units, thereby pruning the decision space. However, when the characteristics of a test instance deviate from the assumptions embedded in these priority models, the solution quality of these algorithms may degrade significantly.
LR decomposes the problem into independent subproblems by relaxing the power balance constraints. Although this decomposition reduces computational complexity, it inevitably discards critical information regarding inter-unit coordination and temporal coupling.
In contrast, the proposed unified solution framework makes no use of domain-specific knowledge or assumptions about individual instances. The framework treats all problem instances uniformly, thereby delivering consistently stable and reliable performance across the majority of test instances.

An advantage of the proposed unified solution framework lies in its theoretical completeness. Given sufficient computational time, the underlying SAT solver is guaranteed to explore the entire solution space and converge to the global optimum.
This theoretical property provides a crucial perspective for interpreting the experimental results. While the proposed framework outperforms existing methods on the majority of test instances (46/54), it does not achieve the best solution on the remaining 7 instances. Our analysis reveals that these exceptions are not due to inherent limitations in the generalizability of the framework, but rather stem from the strict computational time limits imposed in our experiments. On these test instances where the best solution was not achieved, the solver reached the time cutoff before proving optimality, thus returning the best solution found so far. This observation suggests that with relaxed time constraints, the framework would likely close the gap or surpass the current best solutions on these challenging instances as well. Consequently, the observed performance gap in a few test instances reflects a trade-off between computational efficiency and solution exactness, rather than a failure of the framework itself.

In summary, compared with existing UC algorithms, including metaheuristic algorithms, heuristic methods based on power system domain knowledge, and commonly used mixed integer programming approaches, the proposed unified solution framework achieves better solution quality on both the classical unit commitment problem and its variant with ramping constraints. This result demonstrates the generalizability of the proposed framework across different UC variants.

\section{Conclusion}\label{sec:conclusion}

In this paper, to addresse the generalizability limitation of existing UC solution methods caused by the tight coupling between specialized algorithms and the mathematical models of specific UC variants, we propose a SAT-based reduction method and a unified solution framework that decouples the solving algorithm from the problem formulation.
The framework uniformly reduces any UC variant into a standard SAT instance solved by a single standard SAT solver, eliminating the need for variant-specific algorithm redesign.

Experiments on the classical UC and its variant with ramping constraints show that the proposed framework achieves better solution quality and stronger generalizability than specialized algorithms, including metaheuristic, priority-list, Lagrangian-relaxation, and mixed integer programming methods.
These results confirm that a single algorithm decoupled from the problem model can uniformly solve heterogeneous UC variants, and that new variants can be handled by adjusting the input model while reusing the same reduction pipeline and solver.
Future work will extend the framework to a broader range of industrial problems with multiple variants and improve the scalability of the framework on large-scale instances, with the goal of validating and strengthening its generality beyond the UC family.

\bibliographystyle{IEEEtran}
\bibliography{main-ref}

\begin{thebibliography}{10}
\providecommand{\url}[1]{#1}
\csname url@samestyle\endcsname
\providecommand{\newblock}{\relax}
\providecommand{\bibinfo}[2]{#2}
\providecommand{\BIBentrySTDinterwordspacing}{\spaceskip=0pt\relax}
\providecommand{\BIBentryALTinterwordstretchfactor}{4}
\providecommand{\BIBentryALTinterwordspacing}{\spaceskip=\fontdimen2\font plus
\BIBentryALTinterwordstretchfactor\fontdimen3\font minus \fontdimen4\font\relax}
\providecommand{\BIBforeignlanguage}[2]{{%
\expandafter\ifx\csname l@#1\endcsname\relax
\typeout{** WARNING: IEEEtran.bst: No hyphenation pattern has been}%
\typeout{** loaded for the language `#1'. Using the pattern for}%
\typeout{** the default language instead.}%
\else
\language=\csname l@#1\endcsname
\fi
#2}}
\providecommand{\BIBdecl}{\relax}
\BIBdecl

\bibitem{kerrUnitCommitment1966}
R.~Kerr, J.~Scheidt, A.~Fontanna, and J.~Wiley, ``Unit commitment,'' \emph{IEEE Transactions on Power Apparatus and Systems}, vol. PAS-85, no.~5, pp. 417--421, 1966.

\bibitem{Hong:2026kxc}
Y.-Y. Hong and D.~J.~D. Lopez, ``On {{Revolutionizing Power Systems}}: {{Suitability}} of {{Quantum Computing-Ready Pipelines}} for {{Stochastic Unit Commitment Optimization}},'' \emph{IEEE Transactions on Industrial Informatics}, vol.~22, no.~10, pp. 9049--9060, 2026.

\bibitem{huScalableControlLargeScale2025}
X.~Hu, C.~Peng, H.~Shen, and E.~Tian, ``Scalable control of large-scale {{DC}} microgrids: A novel line-independent and mode-related scheme,'' \emph{IEEE Transactions on Cybernetics}, vol.~55, no.~7, pp. 3296--3305, 2025.

\bibitem{yangBurdenHydropowerUnits2018}
W.~Yang, P.~Norrlund, L.~Saarinen, A.~Witt, B.~Smith, J.~Yang, and U.~Lundin, ``Burden on hydropower units for short-term balancing of renewable power systems,'' \emph{Nature Communications}, vol.~9, no.~1, p. 2633, 2018.

\bibitem{guoGridIntegrationFeasibility2023}
X.~Guo, X.~Chen, X.~Chen, P.~Sherman, J.~Wen, and M.~McElroy, ``Grid integration feasibility and investment planning of offshore wind power under carbon-neutral transition in {{China}},'' \emph{Nature Communications}, vol.~14, no.~1, p. 2447, 2023.

\bibitem{luCombinedSolarPower2021}
X.~Lu, S.~Chen, C.~P. Nielsen, C.~Zhang, J.~Li, H.~Xu, Y.~Wu, S.~Wang, F.~Song, C.~Wei, K.~He, M.~B. McElroy, and J.~Hao, ``Combined solar power and storage as cost-competitive and grid-compatible supply for {{China}}'s future carbon-neutral electricity system,'' \emph{Proceedings of the National Academy of Sciences}, vol. 118, no.~42, p. e2103471118, 2021.

\bibitem{mccardleCollaborationsDriveEnergy2023}
K.~McCardle, ``Collaborations drive energy storage research,'' \emph{Nature Computational Science}, vol.~3, no.~6, pp. 464--466, 2023.

\bibitem{srivastavaNewAggrandizedClass2022}
A.~Srivastava and D.~K. Das, ``A new aggrandized class topper optimization algorithm to solve economic load dispatch problem in a power system,'' \emph{IEEE Transactions on Cybernetics}, vol.~52, no.~6, pp. 4187--4197, 2022.

\bibitem{teoOptimizationFuzzyEnergyManagement2021}
T.~T. Teo, T.~Logenthiran, W.~L. Woo, K.~Abidi, T.~John, N.~S. Wade, D.~M. Greenwood, C.~Patsios, and P.~C. Taylor, ``Optimization of fuzzy energy-management system for grid-connected microgrid using {{NSGA-II}},'' \emph{IEEE Transactions on Cybernetics}, vol.~51, no.~11, pp. 5375--5386, 2021.

\bibitem{wangConditionKnowledgeRepresentation2024}
T.~Wang, J.~Zhao, H.~Leung, and W.~Wang, ``A condition knowledge representation and feedback learning framework for dynamic optimization of integrated energy systems,'' \emph{IEEE Transactions on Cybernetics}, vol.~54, no.~5, pp. 2880--2890, 2024.

\bibitem{aliSolvingFuelBasedUnit2024}
S.~Ali, M.~A. {Al-Betar}, M.~Nasor, and M.~A. Awadallah, ``Solving {{Fuel-Based Unit Commitment Problem Using Improved Binary Bald Eagle Search}},'' \emph{Journal of Bionic Engineering}, vol.~21, no.~6, pp. 3098--3122, 2024.

\bibitem{liuHMAMRLMulticriterionFlexible2026}
X.~Liu, M.~Yu, C.~Yang, H.~Wang, L.~Zhou, and H.~Zhou, ``{{HMAMRL}}: Multicriterion flexible coordinated control for coal-fired power generation systems under wide load operation,'' \emph{IEEE Transactions on Cybernetics}, vol.~56, no.~2, pp. 891--903, 2026.

\bibitem{byeonUnitCommitmentGas2020}
G.~Byeon and P.~Van~Hentenryck, ``Unit {{Commitment With Gas Network Awareness}},'' \emph{IEEE Transactions on Power Systems}, vol.~35, no.~2, pp. 1327--1339, 2020.

\bibitem{wangRobustRiskConstrainedUnit2017}
C.~Wang, F.~Liu, J.~Wang, F.~Qiu, W.~Wei, S.~Mei, and S.~Lei, ``Robust {{Risk-Constrained Unit Commitment With Large-Scale Wind Generation}}: {{An Adjustable Uncertainty Set Approach}},'' \emph{IEEE Transactions on Power Systems}, vol.~32, no.~1, pp. 723--733, 2017.

\bibitem{duOperationHighRenewable2019}
E.~Du, N.~Zhang, B.-M. Hodge, Q.~Wang, Z.~Lu, C.~Kang, B.~Kroposki, and Q.~Xia, ``Operation of a high renewable penetrated power system with {{CSP}} plants: A look-ahead stochastic unit commitment model,'' \emph{IEEE Transactions on Power Systems}, vol.~34, no.~1, pp. 140--151, 2019.

\bibitem{zhangMultipleDistributedPVs2025}
B.~Zhang, C.~Dou, D.~Yue, J.~H. Park, X.~Xie, D.~Yuan, and Z.~Zhang, ``Multiple distributed {{PVs}} participating in active power support under resource aggregation and data communication congestion,'' \emph{IEEE Transactions on Cybernetics}, vol.~55, no.~12, pp. 5752--5765, 2025.

\bibitem{dongDatadrivenCostBudget2025}
H.~Dong, L.~Wu, J.~Zhu, S.~Li, Z.~Liang, H.~Yang, and C.-y. Chung, ``A data-driven cost budget satisficing model for unit commitment under solar power uncertainty,'' \emph{IEEE Transactions on Power Systems}, vol.~40, no.~5, pp. 4063--4080, 2025.

\bibitem{jainOptimizedUnitCommitment2025}
S.~Jain and N.~Kanwar, ``Optimized unit commitment for peak load management with solar {{PV}} and storage under load uncertainty,'' \emph{Scientific Reports}, vol.~15, no.~1, p. 19819, 2025.

\bibitem{zhangTwostageStochasticUnit2024}
X.~Zhang, T.~Ding, H.~Zhang, Z.~Zeng, P.~Siano, and M.~Shahidehpour, ``A two-stage stochastic unit commitment with mixed-integer recourses for nuclear power plants to accommodate renewable energy,'' \emph{IEEE Transactions on Sustainable Energy}, vol.~15, no.~2, pp. 859--870, 2024.

\bibitem{liDistributed$Q$LearningBased2020}
F.~Li, J.~Qin, and W.~X. Zheng, ``Distributed ${{Q}}$-learning-based online optimization algorithm for unit commitment and dispatch in smart grid,'' \emph{IEEE Transactions on Cybernetics}, vol.~50, no.~9, pp. 4146--4156, 2020.

\bibitem{correa-posadaDynamicRampingModel2017}
C.~M. {Correa-Posada}, G.~{Morales-Espana}, P.~Duenas, and P.~{Sanchez-Martin}, ``Dynamic ramping model including intraperiod ramp-rate changes in unit commitment,'' \emph{IEEE Transactions on Sustainable Energy}, vol.~8, no.~1, pp. 43--50, 2017.

\bibitem{jinDataDrivenLookAheadUnit2019}
Z.~Jin, K.~Pan, L.~Fan, and T.~Ding, ``Data-{{Driven Look-Ahead Unit Commitment Considering Forbidden Zones}} and {{Dynamic Ramping Rates}},'' \emph{IEEE Transactions on Industrial Informatics}, vol.~15, no.~6, pp. 3267--3276, 2019.

\bibitem{liuAccurateModelingConfiguration2020}
Y.~Liu, L.~Wu, J.~Li, Y.~Chen, and F.~Wang, ``Towards {{Accurate Modeling}} on {{Configuration Transitions}} and {{Dynamic Ramping}} of {{Combined-Cycle Units}} in {{UC Problems}},'' \emph{IEEE Transactions on Power Systems}, vol.~35, no.~3, pp. 2200--2211, 2020.

\bibitem{elsayedThreestagePriorityList2021}
A.~M. Elsayed, A.~M. Maklad, and S.~M. Farrag, ``Three-stage priority list unit commitment for large-scale power systems considering ramp rate constraints,'' \emph{IEEE Canadian Journal of Electrical and Computer Engineering}, vol.~44, no.~3, pp. 329--342, 2021.

\bibitem{faizahDynamicPriorityApproach2024}
F.~Faizah, R.~I. Sudjoko, M.~Syai'in, and A.~Soeprijanto, ``Dynamic priority approach for unit commitment scheduling solution,'' \emph{2024 {{International Electronics Symposium}} ({{IES}})}, pp. 114--119, 2024.

\bibitem{karabasExactSolutionMethod2023}
T.~Karaba{\c s} and S.~Meral, ``An exact solution method and a genetic algorithm-based approach for the unit commitment problem in conventional power generation systems,'' \emph{Computers \& Industrial Engineering}, vol. 176, p. 108876, 2023.

\bibitem{sunLagrangianDecompositionApproach2016}
Y.~Sun, Z.~Li, W.~Tian, and M.~Shahidehpour, ``A lagrangian decomposition approach to energy storage transportation scheduling in power systems,'' \emph{IEEE Transactions on Power Systems}, vol.~31, no.~6, pp. 4348--4356, 2016.

\bibitem{xavierTransmissionConstraintFiltering2019}
A.~S. Xavier, F.~Qiu, F.~Wang, and P.~R. Thimmapuram, ``Transmission constraint filtering in large-scale security-constrained unit commitment,'' \emph{IEEE Transactions on Power Systems}, vol.~34, no.~3, pp. 2457--2460, 2019.

\bibitem{ahmadiSecurityConstrainedUnitCommitment2019}
A.~Ahmadi, A.~E. Nezhad, and B.~Hredzak, ``Security-{{Constrained Unit Commitment}} in {{Presence}} of {{Lithium-Ion Battery Storage Units Using Information-Gap Decision Theory}},'' \emph{IEEE Transactions on Industrial Informatics}, vol.~15, no.~1, pp. 148--157, 2019.

\bibitem{gaoInternallyInducedBranchandcut2022}
Q.~Gao, Z.~Yang, W.~Yin, W.~Li, and J.~Yu, ``Internally induced branch-and-cut acceleration for unit commitment based on improvement of upper bound,'' \emph{IEEE Transactions on Power Systems}, vol.~37, no.~3, pp. 2455--2458, 2022.

\bibitem{luEvolutionaryFractionalOrderExtended2025}
K.-D. Lu, L.~Zhou, and Z.-G. Wu, ``Evolutionary fractional-order extended kalman filter of cyber-physical power systems,'' \emph{IEEE Transactions on Cybernetics}, vol.~55, no.~3, pp. 1395--1408, 2025.

\bibitem{wangScalableAdaptiveFuzzy2026}
A.~Wang, M.~Fei, D.~Du, D.~Wu, C.~Peng, and K.~Li, ``Scalable adaptive fuzzy voltage control for {{DC}} microgrids with plug-and-play operations and constant power loads,'' \emph{IEEE Transactions on Cybernetics}, 2026.

\bibitem{quLinearizationMethodLargescale2024}
M.~Qu, T.~Ding, C.~Mu, X.~Zhang, K.~Pan, and M.~Shahidehpour, ``Linearization method for large-scale hydro-thermal security-constrained unit commitment,'' \emph{IEEE Transactions on Automation Science and Engineering}, vol.~21, no.~2, pp. 1754--1766, 2024.

\bibitem{zhangIncorporatingFrequencySecurity2026a}
R.~Zhang, Y.~Chen, K.~Yan, Z.~Li, T.~Jiang, G.~Li, and X.~Li, ``Incorporating frequency security constraints in two-stage robust unit commitment of integrated transmission and distribution system,'' \emph{IEEE Transactions on Sustainable Energy}, vol.~17, no.~1, pp. 70--86, 2026.

\bibitem{longACNetworkConstrainedUnit2025}
J.~Long, Z.~Yang, Y.~Liu, M.~Xiang, and J.~Yu, ``{{AC}} network-constrained unit commitment based on adaptive linear power flow model,'' \emph{IEEE Transactions on Power Systems}, vol.~40, no.~2, pp. 1360--1373, 2025.

\bibitem{zhaoDistributedResilientSecondary2025}
D.~Zhao, Y.~Li, X.~Cao, Y.~Sun, and C.~Zhang, ``Distributed {{Resilient Secondary Control Strategy Considering Economic Dispatch}} for {{DC Microgrids}}: {{A Dynamic Event-Triggered Mechanism}},'' \emph{IEEE Transactions on Cybernetics}, vol.~55, no.~11, pp. 5250--5263, 2025.

\bibitem{duHighefficiencyNetworkconstrainedClustered2019}
E.~Du, N.~Zhang, C.~Kang, and Q.~Xia, ``A high-efficiency network-constrained clustered unit commitment model for power system planning studies,'' \emph{IEEE Transactions on Power Systems}, vol.~34, no.~4, pp. 2498--2508, 2019.

\bibitem{wuAcceleratingNCUCBinary2016}
L.~Wu, ``Accelerating {{NCUC Via Binary Variable-Based Locally Ideal Formulation}} and {{Dynamic Global Cuts}},'' \emph{IEEE Transactions on Power Systems}, vol.~31, no.~5, pp. 4097--4107, 2016.

\bibitem{chenEfficientMILPApproximation2016}
Y.~Chen, F.~Liu, B.~Liu, W.~Wei, and S.~Mei, ``An efficient {{MILP}} approximation for the hydro-thermal unit commitment,'' \emph{IEEE Transactions on Power Systems}, vol.~31, no.~4, pp. 3318--3319, 2016.

\bibitem{laoImplementationUnitCommitment2022}
S.~Lao, S.~Premrudeepreechacharn, and K.~Ngamsanroj, ``Implementation unit commitment problems for the hydropower plant scheduling use a priority method,'' \emph{2022 {{IEEE}} 31st {{International Symposium}} on {{Industrial Electronics}} ({{ISIE}})}, pp. 865--869, 2022.

\bibitem{zhuTransferbasedApproximateDynamic2024}
J.~Zhu, K.~Zeng, J.~Chen, W.~Zhao, W.~Liu, and W.~Zhu, ``Transfer-based approximate dynamic programming for rolling security-constrained unit commitment with uncertainties,'' \emph{Protection and Control of Modern Power Systems}, vol.~9, no.~5, pp. 42--53, 2024.

\bibitem{chuanluoClauseStatesBased2015}
{Chuan Luo}, {Shaowei Cai}, {Kaile Su}, and {Wei Wu}, ``Clause states based configuration checking in local search for satisfiability,'' \emph{IEEE Transactions on Cybernetics}, vol.~45, no.~5, pp. 1028--1041, 2015.

\bibitem{cookComplexityTheoremprovingProcedures1971}
S.~A. Cook, ``The complexity of theorem-proving procedures,'' \emph{Proceedings of the Third Annual {{ACM}} Symposium on {{Theory}} of Computing - {{STOC}} '71}, pp. 151--158, 1971.

\bibitem{janicicURSASystemUniform2012}
P.~Janicic, ``{{URSA}}: {{A System}} for {{Uniform Reduction}} to {{SAT}},'' \emph{Logical Methods in Computer Science}, vol.~8, no.~3, p. 1171, 2012.

\bibitem{semenovTranslationAlgorithmicDescriptions2020}
A.~Semenov, I.~Otpuschennikov, I.~Gribanova, O.~Zaikin, and S.~Kochemazov, ``Translation of {{Algorithmic Descriptions}} of {{Discrete Functions}} to {{SAT}} with {{Applications}} to {{Cryptanalysis Problems}},'' \emph{Logical Methods in Computer Science}, vol.~16, no.~1, p. 4525, 2020.

\bibitem{zhouOptimizingSATEncodings2017}
N.-F. Zhou and H.~Kjellerstrand, ``Optimizing {{SAT}} encodings for arithmetic constraints,'' \emph{Principles and {{Practice}} of {{Constraint Programming}}}, vol. 10416, pp. 671--686, 2017.

\bibitem{aloulSearchTechniquesSATbased2006}
F.~A. Aloul, ``Search techniques for {{SAT-based}} boolean optimization,'' \emph{Journal of the Franklin Institute}, vol. 343, no. 4-5, pp. 436--447, 2006.

\bibitem{vatambetiSynergisticOptimizationUnit2024}
R.~Vatambeti and P.~K. Dhal, ``Synergistic optimization of unit commitment using {{PSO}} and random search,'' \emph{Contemporary Mathematics}, vol.~5, no.~1, pp. 698--710, 2024.

\bibitem{castilloUnitCommitmentProblem2016}
A.~Castillo, C.~Laird, C.~A. {Silva-Monroy}, J.-P. Watson, and R.~P. O'Neill, ``The unit commitment problem with {{AC}} optimal power flow constraints,'' \emph{IEEE Transactions on Power Systems}, vol.~31, no.~6, pp. 4853--4866, 2016.

\bibitem{kambojNovelHybridPSO2016}
V.~K. Kamboj, ``A novel hybrid {{PSO}}--{{GWO}} approach for unit commitment problem,'' \emph{Neural Computing and Applications}, vol.~27, no.~6, pp. 1643--1655, 2016.

\bibitem{borghettiLagrangianHeuristicsBased2002}
A.~Borghetti, A.~Frangioni, F.~Lacalandra, and C.~A. Nucci, ``Lagrangian heuristics based on disaggregated bundle methods for hydrothermal unit commitment,'' \emph{IEEE Power Engineering Review}, vol.~22, no.~12, pp. 60--60, 2002.

\bibitem{heuleSATCompetition2018}
M.~J.~H. Heule, M.~J. Järvisalo, and M.~Suda, ``Proceedings of {{SAT}} competition 2018 : Solver and benchmark descriptions,'' 2018, \url{http://hdl.handle.net/10138/237063}. Accessed 10 Feb 2026.

\bibitem{ZhangBranchBoundMethod2025}
Z.~Zhang, ``A branch and bound method for solving unit commitment problem,'' \emph{Journal of Electrical Engineering}, vol.~13, no.~1, pp. 10--17, 2025.

\bibitem{saranyaSolutionUnitCommitment2019}
S.~Saranya and B.~Saravanan, ``Solution to unit commitment using lagrange relaxation with whale optimization method,'' \emph{2019 {{Innovations}} in {{Power}} and {{Advanced Computing Technologies}} (i-{{PACT}})}, pp. 1--6, 2019.

\bibitem{huangCorrelationbasedDynamicAllocation2024}
H.~Huang, Y.~Xu, Y.~Xiang, and Z.~Hao, ``Correlation-based dynamic allocation scheme of fitness evaluations for constrained evolutionary optimization,'' \emph{IEEE Transactions on Evolutionary Computation}, vol.~28, no.~5, pp. 1250--1264, 2024.

\bibitem{yangLocaldiversityEvaluationAssignment2023}
S.~Yang, H.~Huang, F.~Luo, Y.~Xu, and Z.~Hao, ``Local-diversity evaluation assignment strategy for decomposition-based multiobjective evolutionary algorithm,'' \emph{IEEE Transactions on Systems, Man, and Cybernetics: Systems}, vol.~53, no.~3, pp. 1697--1709, 2023.

\bibitem{IBMILOGCPLEXOptimizationStudio2024}
{{IBM}}, ``{{IBM ILOG CPLEX Optimization Studio}} 22,'' 2024, \url{https://www.ibm.com/products/ilog-cplex-optimization-studio}. Accessed 15 July 2025.

\end{thebibliography}

\end{document}